# Mars Sample Recovery Helicopter Aerial Mode Commander Design Using Executable Models

**Laura Jones-Wilson**
Jet Propulsion Laboratory, California Institute of Technology
4800 Oak Grove Dr.
Pasadena, CA 91109
Laura.L.Jones@jpl.nasa.gov

**Fatma Karagoz**
Jet Propulsion Laboratory, California Institute of Technology
4800 Oak Grove Dr.
Pasadena, CA 91109
fatma.k.karsten@jpl.nasa.gov

**Maximilian Vierlboeck**
Jet Propulsion Laboratory, California Institute of Technology
4800 Oak Grove Dr.
Pasadena, CA 91109
maximilian.vierlboeck@jpl.nasa.gov

**Thomas Boyer-Chammard**
Jet Propulsion Laboratory, California Institute of Technology
4800 Oak Grove Dr.
Pasadena, CA 91109
thomas.boyer.chammard@jpl.nasa.gov

**Vinod Gehlot**
Jet Propulsion Laboratory, California Institute of Technology
4800 Oak Grove Dr.
Pasadena, CA 91109
vinod.p.gehlot@jpl.nasa.gov

**Håvard Grip**
Jet Propulsion Laboratory, California Institute of Technology
4800 Oak Grove Dr.
Pasadena, CA 91109
havard.f.grip@jpl.nasa.gov

*Abstract*— As our knowledge of other worlds deepens, space exploration projects must support increasingly sophisticated payloads, tolerate more extreme environments, and generally manage more engineering complexity to push the envelope. The growing technical demands of these projects pose unique challenges to systems engineering in managing risk as nuanced designs are communicated across diverse teams. To meet this challenge, systems engineering practices must evolve to better support intentional architecture and the communication/ infusion of that architecture into downstream development, integration, and testing. For instance, the needed behaviors of a system must be articulated early in the project lifecycle, and those behaviors drive design and testing (and cost). Using formalized models of the behaviors, we can catch errors in design early, better communicate requirements to developers, provide a framework for the implementation of the behavior, and feed-forward the behavioral needs into test cases and operations. The formalized modeling approach preserves the design intent and constraints in ways that are more robust and potentially time-efficient than more traditional documentation and interface handoffs often are. This paper explores a case study of the use of behavioral modeling in a mode commander for aerial flight of a Sample Recovery Helicopter (SRH) mission concept that was designed to retrieve and deliver sample tubes to the Sample Retrieval Lander (SRL) as a part of the planned Mars Sample Return (MSR) mission.

In particular, we describe the application of model-based systems engineering with executable system modeling for describing the behavior of the Aerial Mode Commander of the helicopter. Descriptive and behavioral models are used in this context to: 1) provide consistency and tighter integration between flight system description and flight software implementation, 2) enable early detection of any unexpected/undesired behavior by tracking the system states for different mission scenarios, 3) qualify the comprehensiveness of the behavioral descriptions via simulations. The execution of the system model enabled the dynamic evaluation of the system behavior, qualification of mode commander design, and validation of scenarios and test cases.

Formal methods are leveraged to represent the system and validate its behavior by visualizing and simulating different mission scenarios and test cases. Interactions between aerial mode commanders are explicitly defined using SysML, which provides rigor and flexibility to adequately capture the level of detail needed to reduce ambiguities in design, ultimately leading to a more consistent and reliable system. Model patterns are constructed to address specific flight computer interactions and fault behaviors. Early validation of system behavior offered resources for a streamlined transition from design to implementation, leading to more robust and efficient flight software that performs as expected across various mission scenarios.

## Table of Contents

# 1. MOTIVATION AND OVERVIEW

Strong system architecture definition has long been recognized as a foundational part of NASA missions [1] and much of our systems engineering processes are devoted to preserving and communicating architectural intent across design phases – all the way to validation, the ultimate check of a design cycle output to its intended function. How effectively the early conceptual intent is flowed through the later project phases has a direct impact on implementation success – in terms of cost, schedule, and technical integrity. So as the systems engineering discipline moves toward a more model-driven (rather than document-driven) paradigm, it is important that we ensure these new techniques still serve to promote a shared understanding of the system across project phase and diverse teams at multiple project levels.

Model-based systems engineering approaches rely on international modeling standards such as the Object Management Group (OMG) Systems Modeling Language (SysML) to formalize and codify the integration of requirements, design, analysis, and validation to ensure that complex systems exhibit the desired behavior. As associated SysML tools/processes have matured and evolved, researchers have started examining how these techniques might be leveraged to improve the connection from early-phase architecting to development. [2] [3] [4] [5] For instance, SysML can offer a more rigorous structure by which to define and develop canonical representational diagrams that are often used to describe and communicate design, and can be used to autocode logic/stubs for some software. [6]

However, the benefits of this approach can go deeper than merely improving "flat" representational diagrams. [7] Executable SysML models are integrated with a set of analysis patterns that leverage activities and state machines directly, which means they can simulate the dynamic behavior of the system, capturing how it transitions between states and executes specific activities under various conditions. If a standard "flat" diagram is a blueprint for a design, the executable model is the running prototype model that can be driven in much more flight-like ways. In order to build these types of models, the system engineer must elaborate on the hierarchical relationships in the system, essentially defining a much deeper architecture than required by typical diagrams. And, for execution to be successful, the model must be semantically correct, so an executable model implies a level of semantic integrity that inherently facilitates the verification of model correctness. These types of models have been studied as a way to more precisely document architectures, conduct trades, and evaluate performance against requirements [7] [8] [9].

As the evidence builds that there are potentially many powerful advantages to this approach, there are still relatively few examples in literature of its use on active projects in their design process (rather than researchers or modelers using a project as a case study that was not implemented by the project itself). Further, the literature often only investigates examples at a high system level, leaving one to wonder whether the benefits found for level 1 or level 2 of a project would hold up when applied at a more detailed subsystem design. These questions become especially important given the heightened awareness of the criticality of software in all aspects of a system design process, and the natural traceability of executable models to software development that may amplify the benefits of this approach for a software-based product.

In this work, we describe the use of formal behavioral modeling (via executable models in SysML) for the Aerial Mobility mode commander for a Mars helicopter flight project called Sample Recovery Helicopter (SRH). The executable models were developed in Cameo System Modeler (CSM) by the Aerial Mobility systems engineer, in concert with the modeling experts, and were used by the project throughout the early design phase of the mission. We will first describe the system being modeled and the unique modeling patterns that were developed to capture actual behavior implementations such as interdependent state machines and hot swap/fault management. These patterns are likely to be relevant to a wide variety of mode commanders. We then describe the benefits of this approach, especially how it promoted a more disciplined architectural definition than graphical models alone might, allowing the architecture to be more thoroughly exercised before its final implementation in code. Although SRH was canceled just after its preliminary design review (PDR), the paper describes expected benefits of automated code generation and verification and validation test case generation enabled by this approach as compared to a traditional (non-model-based) approach. The mode commander described was implemented in a released version of the guidance and control (G&C) code, so a summary of that process and results are also included to round out the discussion of the role that executable models can and did have in this flight project.

# 2. SRH DESCRIPTION

### *A. Mission Description*

The goal of the Mars Sample Return (MSR) campaign being planned jointly by NASA and the European Space Agency is to bring sample tubes collected on the Martian surface to Earth so those samples could be analyzed in terrestrial laboratories by the planetary science community. Some of these samples have already been collected by the Mars 2020 mission with the Perseverance rover. [10] The MSR architectural concept calls for a Sample Retrieval Lander (SRL) that is stationary on the Martian surface while samples are delivered to it by a surface asset. The SRL then packages and launches the cache of samples from the Martian surface. [11] The Sample Recovery Helicopter (SRH) project was tasked with designing a backup vehicle that could recover sample tubes and deliver them to SRL. [12] Although the SRH project was descoped from MSR in spring of 2024 as part of a broader campaign-level rearchitecting, SRH achieved its preliminary design in November 2023 and, in part by heavily leveraging the technology and experience from Ingenuity, [13] had achieved an initial delivery of updated guidance and control (G&C) flight software – to include the mode commander discussed in this work – at the time the project was shuttered.

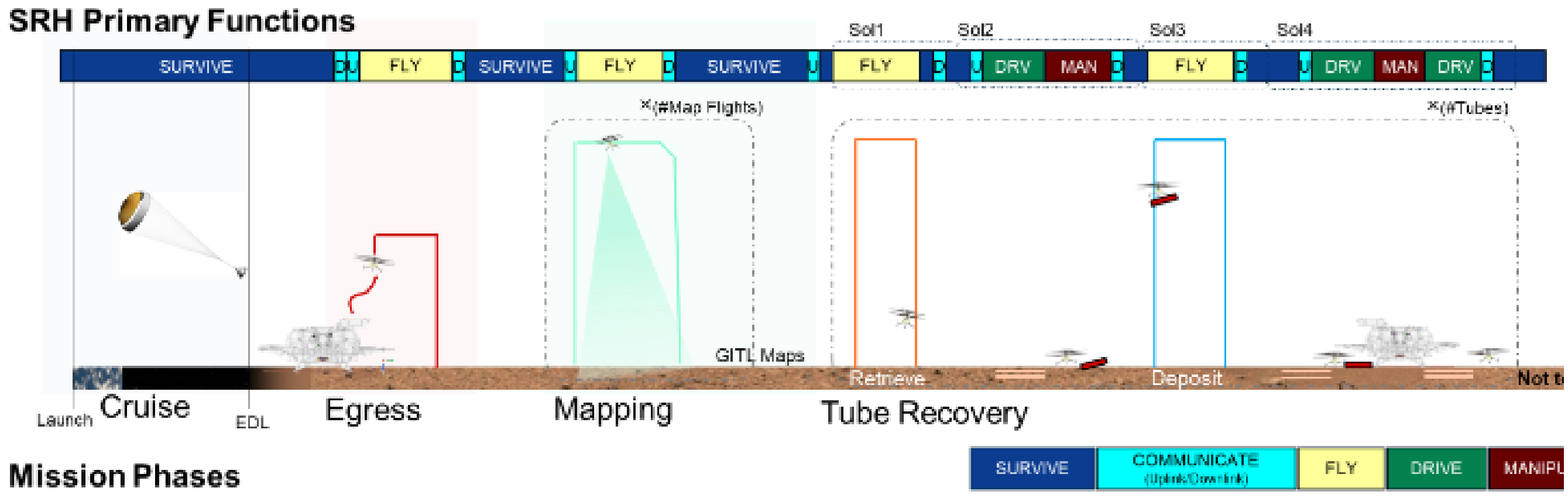

Fig. 2 Sample Recovery Helicopter functions, mission phases, and concept of operations

SRH's mission concept, shown in Fig. 2, called for two identical SRH vehicles brought to the Martian surface as payloads on the SRL. Once landed and configured, each 2.7 kg helicopter would in turn perform a mechanism-boosted egress flight away from the lander. From there, the helicopters perform a set of mapping flights to collect images over the tube depot and lander drop off areas. Once the ground localized the mapping images and reuploaded that information to the helicopters, the tube retrieval phase of the mission begins, with each helicopter taking 4 sols to fly to a given sample tube, collect it, fly back to the lander, and deposit it in the collection area. This process repeats for the number of tubes being collected. This concept was often summarized by the functions "Fly" "Drive" "Manipulate" as they were separate activities being managed by different subsystems and never performed in parallel.

SRH drew heavily from the heritage achieved by the Ingenuity Mars helicopter, in part due to time pressure to meet the SRL launch date at the time. Thus, SRH's helicopters were also coaxial single-rotor vehicles with blade designs, rotor systems, and flight dynamics to similar to Ingenuity. We also retained Ingenuity's approach to flight controls: a collective-cyclic actuation with a fixed rpm during a given flight. [14] [15] [16] [17] [18] While Ingenuity was not designed to have the ability to drive or manipulate samples, it successfully demonstrated the technology to survive entry-descent-and-landing (EDL) and operate on the Martian surface, communicate with another asset on Mars, and fly using visual odometry (VO) to track features on the surface. [13] [19] [20]

### B. *SRH Functional Approach and Responsibilities*

Along with these technical capabilities, the SRH concept also drew from Ingenuity's philosophy and approach to the engineering design. For instance, as with Ingenuity, SRH chose to split major vehicle functions between sub-teams and space those functions in time to allow for crisp, easily-verified interfaces. Following this philosophy, flights are never combined with a drive or manipulate action, allowing the "fly" function to be designed and validated directly by the Aerial Mobility subsystem responsible for implementing that function. The "fly" interface with the "drive" or "manipulate" functions were thus limited to requirements such as those on landing delivery accuracy rather than deeply interdependent software. Further, decision-making for the vehicle was largely consolidated to the software executing the function currently at hand, limiting complex interactions across systems. For example, when the vehicle is executing on a flight sequence, the Aerial software was fully in charge of the vehicle, with other functions (such as fault monitors across the system) setting flags that the Aerial software can choose to react to depending on the state of the vehicle. This idea of "separable functions" allowed the SRH team to retain key elements of the Ingenuity approach to development that proved to be so successful.

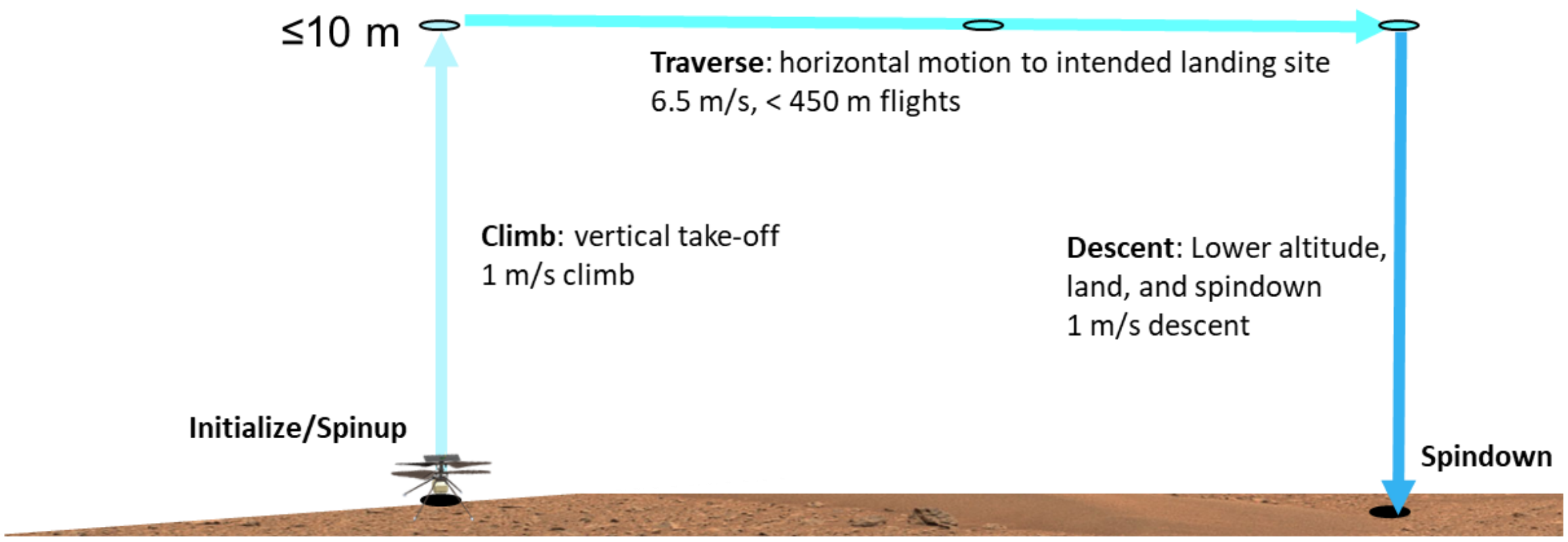

Fig. 1 Anatomy of a flight sequence for SRH

To that end, the Aerial Mobility team and the Avionics teams together handle the Flight SoftWare (FSW) development for the vehicle, but the system-level FSW calls the Aerial Mobility (or guidance, navigation, and control, or GNC) flight software when the vehicle is set to perform a flight. Avionics develops the infrastruture components, including drivers for sensors and actuators, software interfaces, and operational software, among others. The Aerial Mobility team implements the flight controller, navigations filters and estimators, and the Aerial mode commanders required for flight. The Avionics is reponsible for integrating all of the software components using the F Prime [21] FSW framework to produce the FSW binaries that run on the FC and the Nav flight computers. [22]

### C. *Aerial Flight*

This paper focuses on the modeling of the SRH Aerial mode commander, which executes the "Fly" function for the helicopter. While the SRH mapping flights had a very similar structure and design to Ingenuity's flights, the egress and sample tube retrieval flights called for new functionality. The SRH team thus started with the Ingenuity Aerial mode commander [23] and augmented the design to serve SRH's new needs.

All of SRH's flights shared a common set of phases as shown in Fig. 1: initialization, rotor spinup, climb, traverse, and descent. Within this overall framework, different flight types could use different sensors or algorithms, different transition triggers, or unique modes as needed. Roughly the phases correspond to:

- *Initialization*: evaluates the health of sensors and starts up the estimator
- *Spinup*; brings the rotor to its maximum planned rotational speed but with the rotor blades pitched in such a way that they do not generate thrust.
- *Climb*: applies thrust until achieving a pre-set target time corresponding to a particular altitude.
- *Traverse*: maintains its cruise altitude (10 m for SRH) and navigates along its planned traverse.
- *Descent*: once above the landing location, descends until a touchdown is detected and the system is able to spin down the rotors

Fault responses are intentionally localized and simple; if there is a fault while not airborne, the vehicle ensures the rotors spin down and return to idle. If the vehicle is already airborne, it ignores faults during especially critical times (such as immediately after takeoff) and attempts to land immediately in modes where landing improves the chances of vehicle survival of the fault.

The simple linear nature of the flight behavior – in that initialization must happen before spinup, which must happen before a climb, etc. – means that the mode commander also follows a linear pattern, where one mode necessarily leads to the next one. Thus, while there are many modes, the mode commander is not as complex as the number of modes might imply in a more traditional spacecraft.

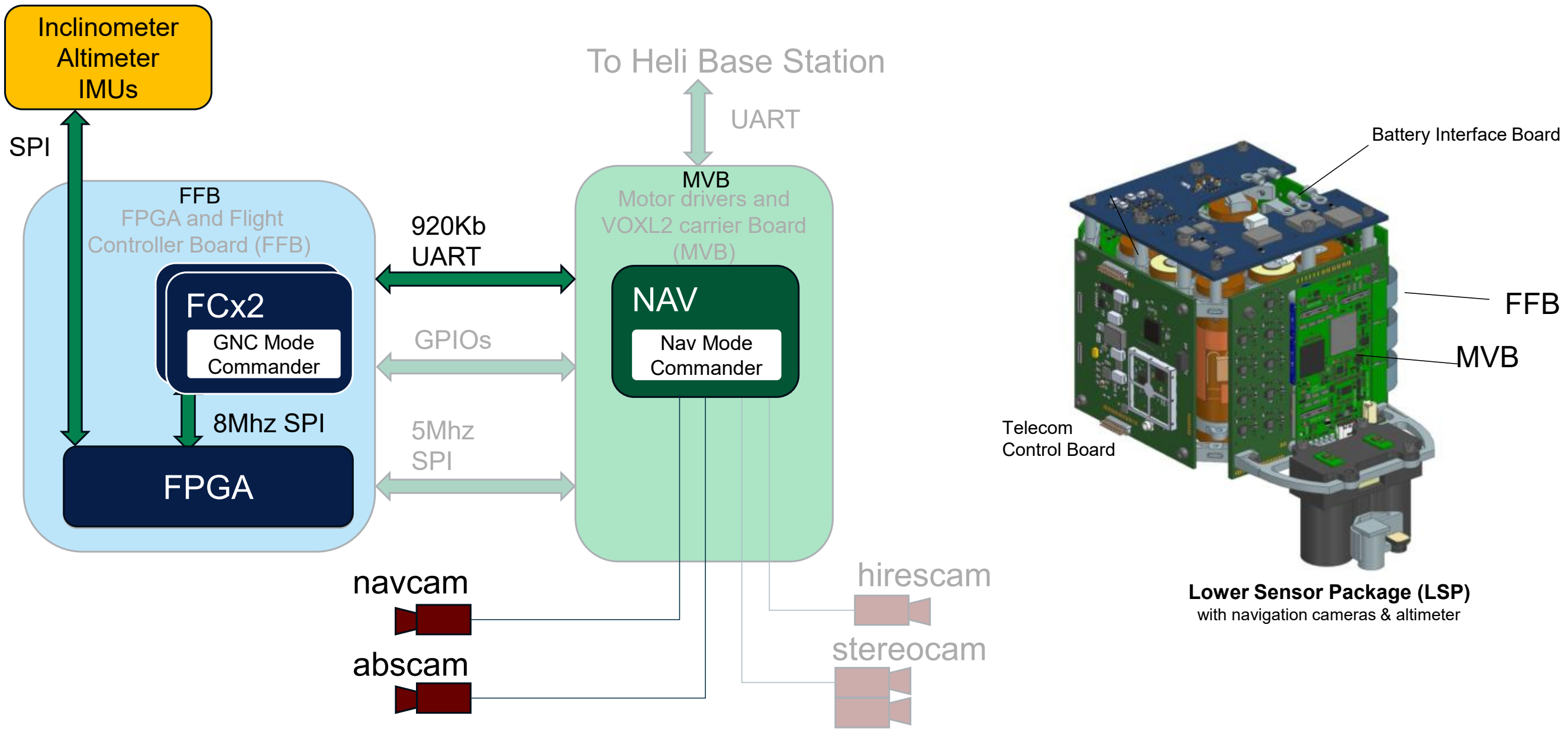

**Fig. 3** Hardware on which the Aerial mode commanders run, including redundant Flight Computers (FCs) and a separate FFB (FPGA and Flight Controller board) and MVB (Motor Driver and VOXL2 Carrier Board)

*D. Aerial Mode Commander Implementation*

Even being linear and relatively sparsely connected, the Aerial mode commander has some complexities because it must coordinate activities across a number of different control system elements: the flight phase, the controller type, the guidance input, and navigation/estimation mode, as well as managing any faults in the system during the flight system. Because of our Ingenuity heritage, we had a relatively mature hardware/avionics design that allowed us to add a richer modeling approach than is common in the early design phase of a project, but the implementation approach did add some layers of complexity and nuance that posed unique challenges to the modeling effort. As seen in Fig. 3, the navigation element of flight is managed on a different board than the flight computer to support the higher processing power needed to interpret the camera imaging data. Thus, the navigation mode commander is housed on a different board than the flight computer mode commander that controls the other elements of the flight. These two mode commanders must be synchronized for the flight to work properly. Further, there are two flight controllers for redundancy. If one board fails, the other one is "hot swapped" in to support the vehicle as it attempts to make an emergency landing. This design has two copies of the GNC mode commander, one on each flight computer, that must be ready to take over in the event of a failure during flight.

A non-executable representation of the Aerial mode commanders is shown in Fig. 4. The "G&C mode commander" (G&C MC), which runs bare-metal on the dual ARM Cortex-R5 — 2X TI Hercules TMS570 — on the FFB serves as the main ringleader for the flight sequence. It dictates which phases of flight the system is in, and orchestrates the controller mode and guidance modes (which are also orthogonal states to the G&C MC). The "Nav mode commander" (NAV MC) runs in userspace in a (preemptive) Linux OS running on the 8-core Qualcomm QRB5165 application processor located on the MVB and dictates which mode the estimator is in. This diagram is a simplified version of a traditional way to communicate how the mode commanders should be designed; by simply drawing out the states and how they are connected.

*E. Aerial Mode Commander Modeling Approach*

When translating this representation into an executable model, the implementation requires that we be able to model synchronized, separate mode commanders, the ability to duplicate signals to move the GNC mode commander on both the prime and backup flight computer to the same modes at the same time, and the ability to "hot swap" to the backup GNC mode commander while retaining the ability of the single nav mode commander to transition properly based on which flight computer (FC) is currently prime. The fact that our design was based on actual code that was flying on Mars allowed our team to explore modeling nuances that were anchored by the actual implementation. This fact provided insights into our modeling approach that strengthened our model fidelity and provided an opportunity to directly compare how our modeling approaches approximated the mature design of very similar software.

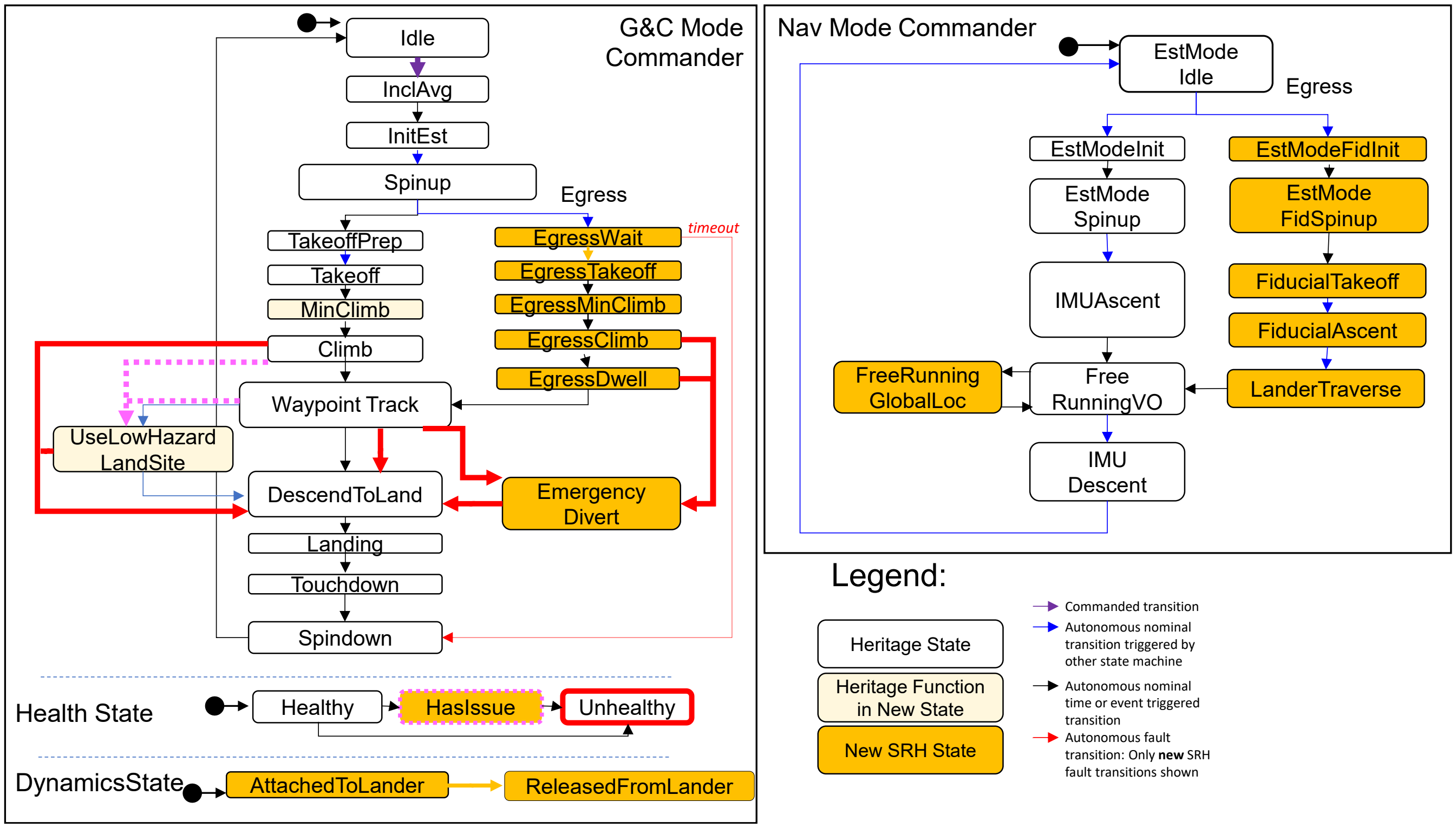

**Fig. 4** SRH Aerial Flight Mode Commander Design in Non-Executable Form, with some modes not shown

SRH chose to explore the use of SysML executable models in the design of its Aerial mode commander because it represented a unique opportunity to study and internalize the Ingenuity architectural intent (by modeling the Ingenuity mode commander first), then ensure that intent was carried across to the SRH design by replicating the structure and order of the Ingenuity design and making SRH-specific changes that fit within it. Because executable models enable the simulation and testing of system behavior, this approach provided a framework for validation and verification of the Aerial behaviors throughout the development lifecycle as alterations were made to suit SRH needs. A snapshot of the executable models is in Fig. 5. It also helped to further refine/formalize/document the mode commander as SRH sought to bring its inherited designs from a technology demonstration risk posture to one more in line with the mission objectives.

## 3. Synchronized Mode Commanders

### *A. G&C MC and NAV MC System*

As described in Section 2.D, SRH's avionics design closely mirrored the structure of Ingenuity's avionics design, with two boards each serving different but critical functions in the helicopter flight. As a result of this hardware design, the aerial mode commander was fundamentally split between two mode commanders that communicate across a UART interface yet must be synchronized for a successful flight. For instance, if the navigation is in the wrong mode to support climbing the control system will not have a correctly formed estimate of its state to react to and the vehicle will not successfully fly. In fact, many state transitions in the two mode commanders are set up to wait for the other mode commander to be in the correct state before advancing, shown as blue arrows in Fig. 4. For instance, the G&C mode commander will not advance to rotor spinup ("Spinup") unless the Nav mode commander is in the estimator mode that supports spinup (EstModeSpinup). Information on the mode of each mode commander are included in a data package shuttled back and forth between the boards by the system flight software function at 500 Hz. One of the most fundamental patterns underlying the Aerial mode commander architecture, then, was the ability to properly capture how these two state machines are dependent on each other for the transitions.

### *B. Interdependent State Machines*

As described in the first section, state machines are an essential part of executable models. In the case of the Sample Recovery Helicopter, multiple state machines were built, as different parts of the system were modeled in alignment with the components described in Section 2. The modeling of these individual components as state machines required consideration of interdependencies, since the individual state machines do not only need to function independently but also rely on each other for inputs, outputs, and conditional behaviors. As a result, the following two dependencies had to be addressed, which necessitated the creation of patterns described below: communication interface and conditional guard dependencies.

Since the components and subsystems within the SRH depend on each other, the state machines used to model the components had to exhibit the same behavior. The dependencies between the components are twofold: signals are sent back and forth between components, and certain state transitions depend on the status of other components, either as a trigger or as a condition to define the target state. The modeling and actual realization require that blocks within the SysML model can read and access the state and status of other blocks and state machines. To achieve this, the interface is created with ports and connectors between mode commander parts that needs to communicate.

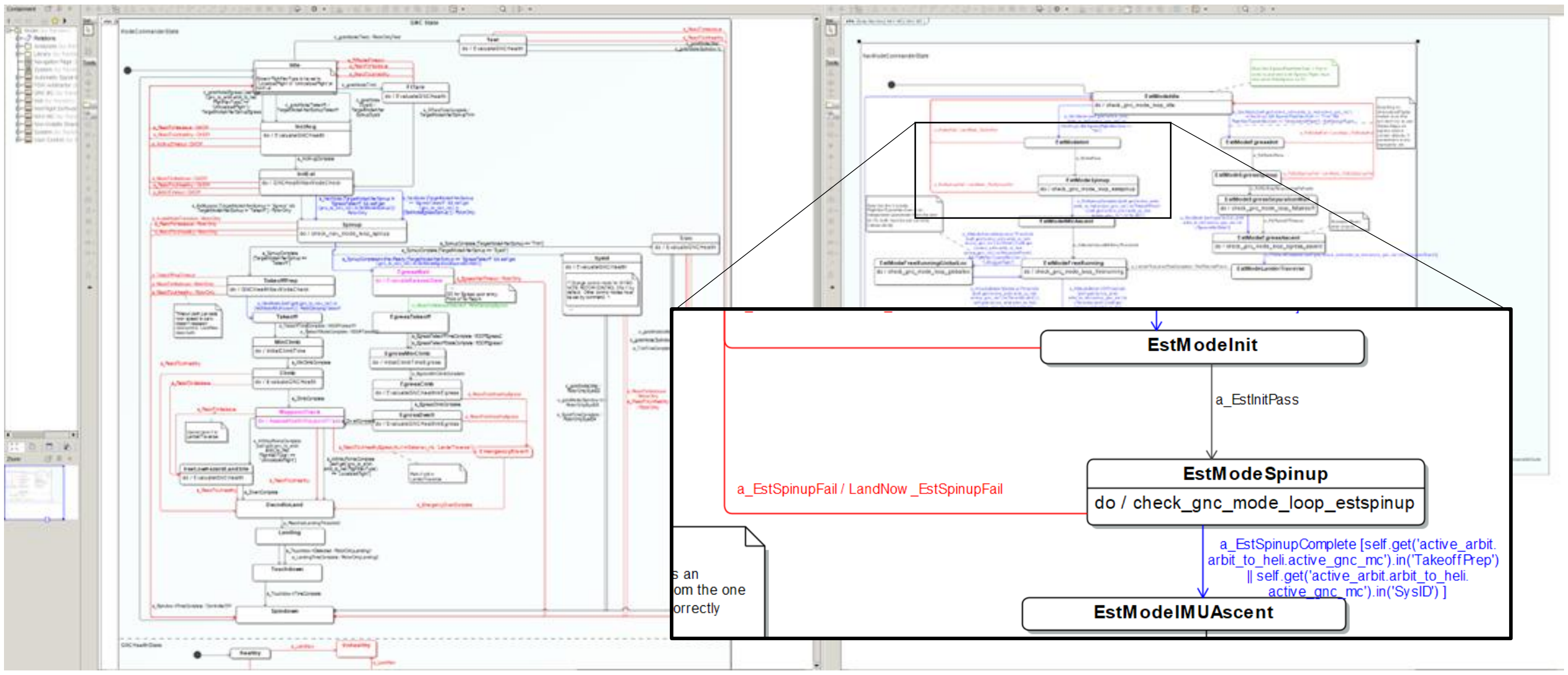

**Fig. 5** A snaphot of the SRH Aerial Flight Mode Commander Design in Executable Form, the GNC MC on the left and the NAV MC on the right, with blue signals correspecting to autonomous nominal transitions triggered by interdependent state machine status and red signals corresponding to fault transitions related to vehicle health status. A zoomed-in image of the transitions for EstModeSpinup are shown, including an "automatic" transition of the initialization "passing" (a_EstInitPass) and an interdependent transition where the other mode commander must be in TakeoffPrep or SysID mode in addition to the estimator spinup being complete (EstSpinupComplete) in order to move to the IMU Ascent mode.

To apply evaluation of guard expressions for interdependent state machines, action languages are used to inject custom logic to guards on transitions and to simulate the conditional expressions in the model. Specifically, ALH (Action Language Helper) APIs [24] are used in Cameo Systems Modeler [25] to query a value from a system parameter and to check whether a particular model component is in a specific state. During a state transition, these guard expressions are evaluated, and transitions occur based on the ALH API responses. Hence, ALH provides a structured way to define conditions and interactions within the state machines, such as evaluating states of another machine or checking values of parameters.

### *C. Pattern Implementation*

Implementation of interdependency is achieved by utilizing a combination of Internal Block Diagrams (IBDs) and Parametric Diagrams. While IBDs establish the communication pathways, parametric diagrams set the rules for these interdependencies. This pattern allows multiple state machines to operate in a coordinated manner, responding dynamically to each other's states and system-wide changes. State transitions needed guards to access the modes/states of other machines. For instance, the following guard below reads the required state from the target machine and only allows a transition if the condition is met:

*“self.get(‘StateMachine1).in(‘StateA)”*

This guard checks if *StateMachine1* is currently in the state *StateA*. Only when this condition is true will the transition in the current state machine be triggered. Similarly, ALH syntax allows for retrieving a structural feature value, which is also utilized in this work to create combined guard conditions that ensure the transition takes place only under desired circumstances. For example:

*“self.get(‘StateMachine1).in(‘StateA) && parameter == 5”*

will evaluate as “*True”* if parameter has a value of 5 and StateMachine1 is in *StateA*.

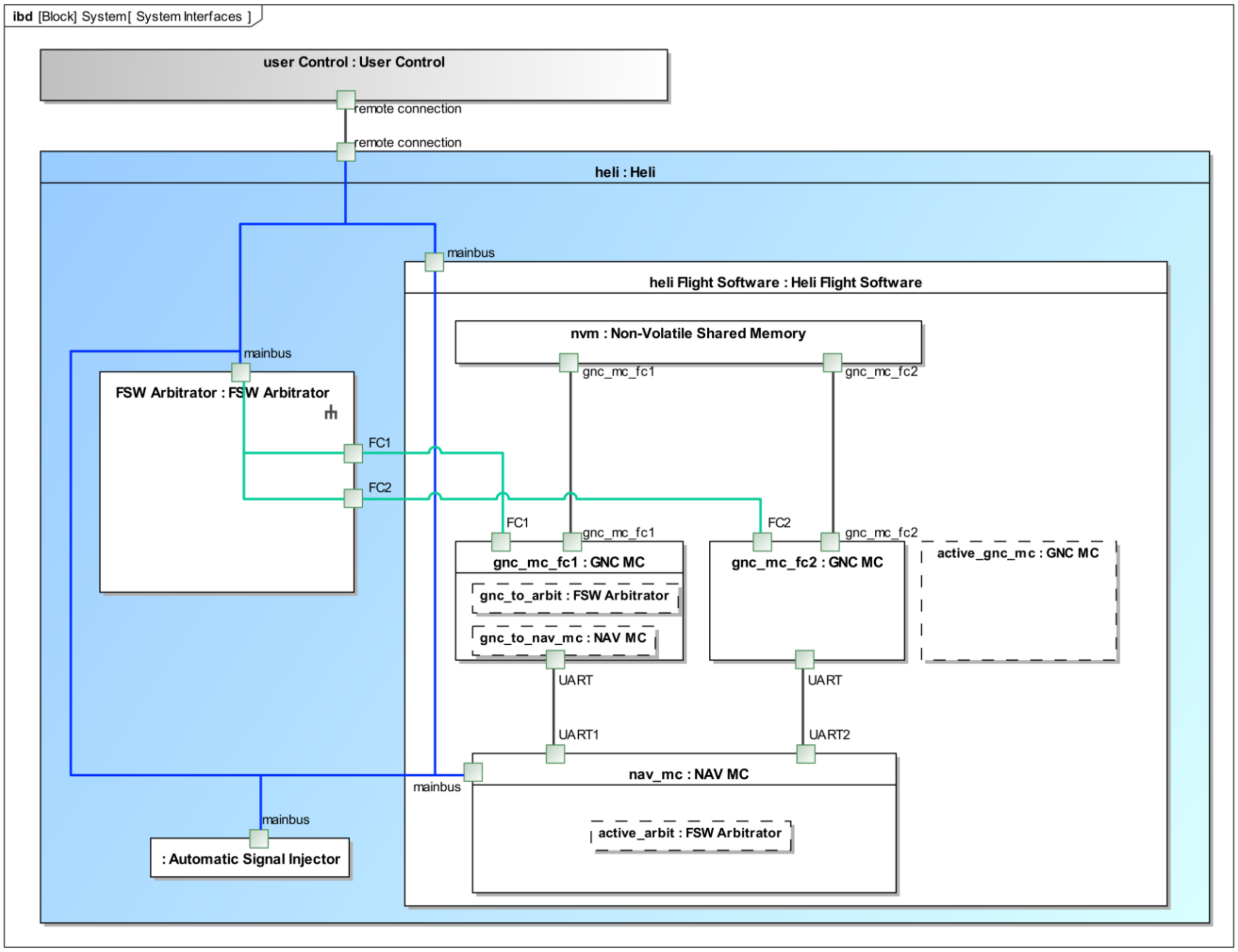

**Fig. 6** System interfaces between system components

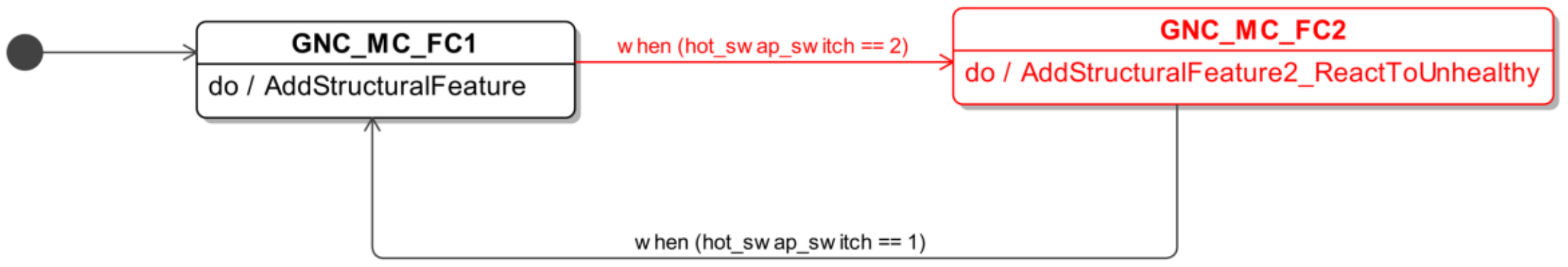

**Fig. 7** Description of the hot-swap pattern in the context of helicopter flight software

The system interfaces that are created to enable the interdependency is described in Fig. 6, where the communication pathways are captured via ports and connectors specifically between NAV and GNC mode commanders. Reference properties are created to accommodate the information access between system components. This provides a flexible means of accessing information across different parts of the system without duplicating or embedding elements directly. For example, *gnc_to_nav_mc* reference property enables access to NAV mode commander's attributes without owning or containing that block. This way, the model can maintain a modular system structure where the mode commanders can still communicate with each other without direct containment. As such, state machines can dynamically evaluate conditions based on real-time data from other system components. Further details regarding additional interfaces and properties will be discussed in the next section.

## 4. Fault Behavior: Hot Swap

### *A. SRH Aerial Fault Behavior*

As helicopters are inherently unstable dynamical systems in flight, reacting to a fault requires very quick action to improve the chance of a successful emergency landing. SRH incorporated a number of fault management features to manage issues during a flight. Much was directly inherited from SRH, where a "LandNow" flag could be set in the Aerial software and as the system transitioned through different G&C modes, it could choose to react to the LandNow flag by aborting its nominal next steps and trying to land as quickly as possible. This was modeled directly in the mode commanders where the "heath" status was an orthogonal state to the G&C mode commander, as shown in Fig. 4. A LandNow signal thrown anywhere in the system would set the health state as "Unhealthy" and as the system progressed through the flight mode, a "health monitor" activity running on the timer corresponding to that board's regular time interval (RTI) rate would see the health status and allow an exit for that state if appropriate.

Another failure management feature of the Ingenuity design that SRH inherited was the robustness to flight computer failure. The helicopter has the ability to hot-swap to a backup processor if a failure in the prime FC is noted. In this design, two flight computers operate in parallel, both running the G&C mode commander, but with only one being active and in command of the vehicle unless a "hot swap" is triggered, and the backup then becomes prime. A small amount of information is stored in non-volatile memory so that in the event the backup processor must take over, it can start in the correct mode with an estimator solution. The primary goal of the system design is to allow the backup FC to resume control of the vehicle within a short number of RTIs, and executable system models with sufficient fidelity are able to probe a design's suitability for achieving that goal without any flight code being written.

### *B. Hot Swap Pattern*

In simple terms, the hot swap pattern we chose can be imagined as running both commanders concurrently from the beginning while one of them is the driving unit, and, after a critical failure event, disconnecting the first unit and immediately switching to the backup unit. This HotSwap pattern required the development and application of two important mechanisms: concurrent state machines/transitions and dynamic dependency targets.

G&C mode commander not only had dependencies (see Interdependent State Machines) on other components, but other components also depended on the G&C MC. For example, the NAV MC executed certain transitions based on the state of the active G&C MC. Since dependencies typically point to a specific component that does not change over time, the problem of dynamic targets and dependencies had to be addressed. To accomplish this, a storage attribute was defined to indicate the identifier of the currently active FC. Based on this identifier, components depending on the state of the active G&C mode commander could query and access the state of the active one, thereby using the correct information for their activities and transitions. This became especially critical after a failure event and an executed HotSwap, as the previously active mode commander had to be considered faulty from that point onwards, despite both having been in sync before the switch.

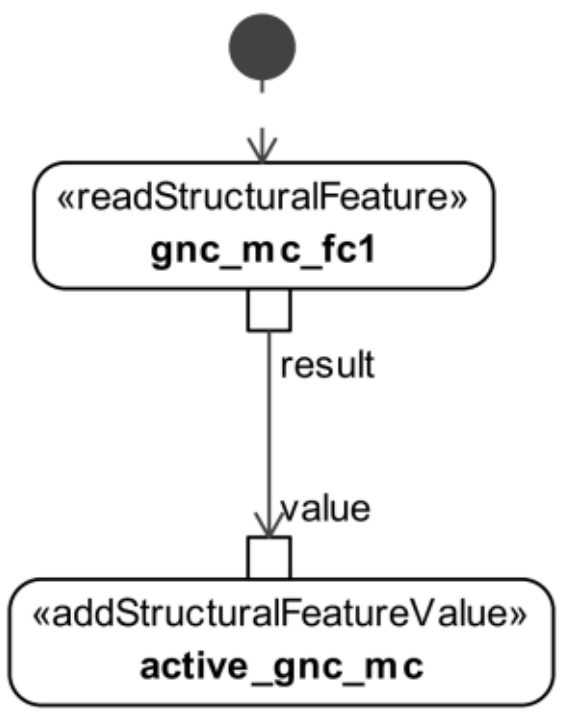

**Fig. 8** Acvitiy assigning current G&C mode commander as active

In order to address these dependencies and maintain a smooth handover between G&C mode commanders, *active_gnc_mc* reference property is created in the model (as shown in Fig. 12) which is able to read and maintain the information from the active GNC mode commander at any given time. This is achieved by the pattern illustrated in Fig. 7, where the helicopter flight software goes through a state change when *hot_swap_switch* value property changes, with a change event trigger. This state change starts the do activity *AddStructuralFeature*, which enables the following:

1. Retrieve the information from the current GNC mode commander via Read Structural Feature action
2. Adding of this information to the *active_gnc_mc* via Add Structural Feature action

Both Read Structural Feature and Add Structural Feature actions are constructs that are used to manipulate and access system properties within the context of activity diagrams or behavior models that facilitate dynamic querying and modifying structural elements of the system. Fig. 8 describes the sequence of usage of these constructs within the *AddStructuralFeature* do activity, depicting a scenario in which the *gnc_mc_fc1* mode commander is assigned as the active mode commander.

Additionally, since both G&C mode commanders are active after boot, concurrent behavior and aligned state transitions were necessary to ensure that the backup commander could take over at any point in time in the state of the failed commander. To achieve concurrent behavior of both commanders, signal duplication was used to relay incoming signals to both instances, while simulation configurations were defined to invoke multiple instances of the same state machine. This setup enabled the modeling of two GNC mode commanders as part properties (see Fig. 6), that remained synchronized as they processed the same signals and executed the same state transitions. This way not only can a single signal be multiplied in one direction, but also signals from multiple sources can be merged in the other direction, mimicking the behavior of traditional communication for distributed senders/receivers. It is worth noting that, depending on the tool, the multiplication of signals may be displayed as consecutive signal flows, due to the fact that some applications only allow for the display of a single token at a time. Nevertheless, the signals are multiplied and subsequently processed in parallel. This behavior is disrupted only if a failure occurs, triggering the initiation of the HotSwap behavior and switching over to the backup commander. With the simulation framework, the states for each commander instance could also be monitored and recorded, allowing for comprehensive testing and troubleshooting.

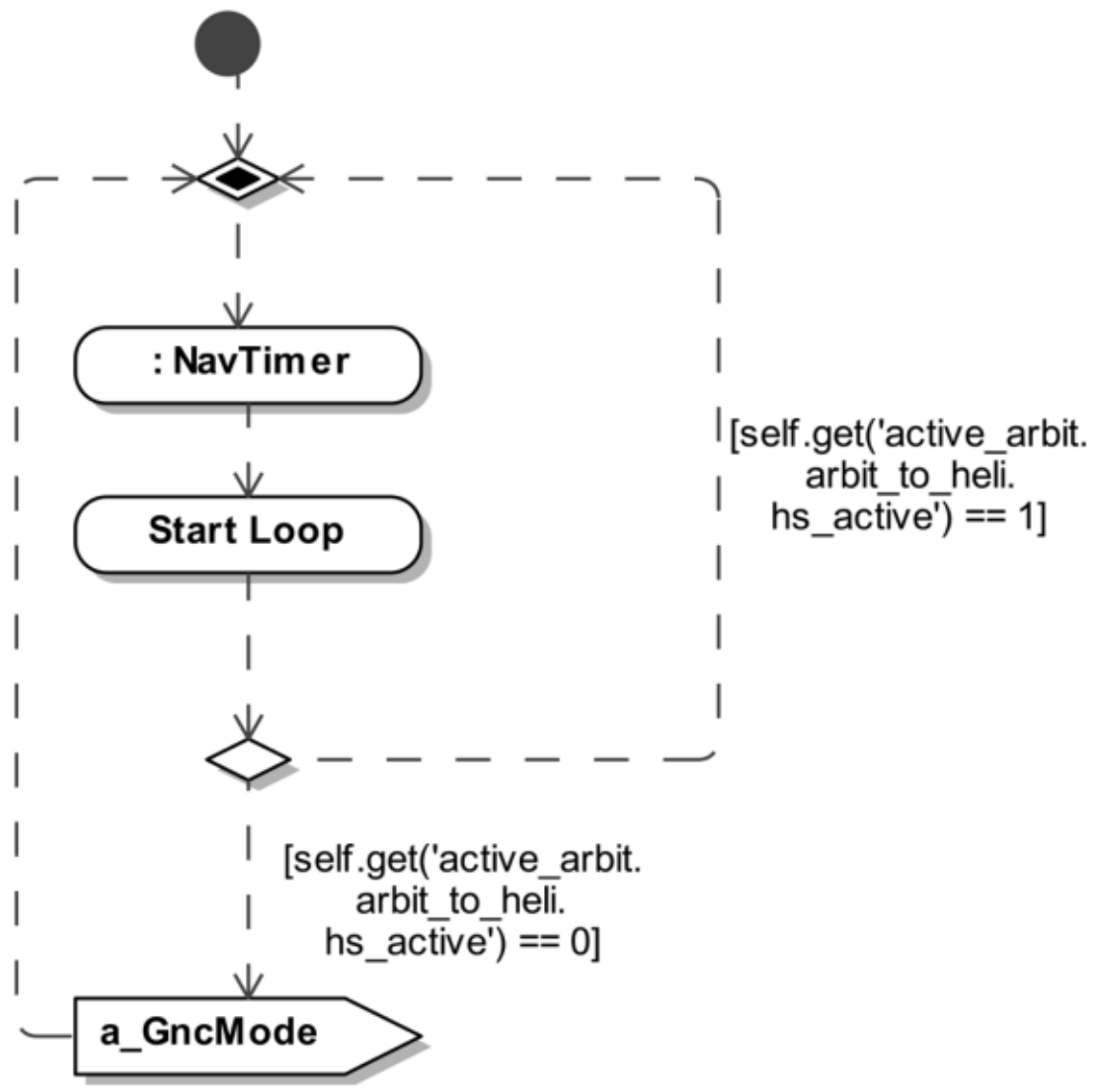

**Fig. 9** Repeated signals with HotSwap guards

The signal duplication is achieved by defining an arbitrator block designed to receive a signal or input and then distribute it to multiple output ports. Fig. 6 illustrates how the signal from the arbitrator block (*User Control*) is routed to both GNC mode commanders (*gnc_mc_fc1* and *gnc_mc_fc2*) through *FSW Arbitrator*.

By managing the interdependencies and synchronization between flight computers, modeling of the HotSwap pattern enabled two key benefits and led to the discovery of potential vulnerabilities in the system design that could be addressed as part of the subsequent system development, stretching all the way to potential hardware implications.

First, the modeling allowed for the evaluation of the boundaries between permanent storage and transient memory. Resulting from the fact that the primary mode commander controlled the behavior until the failover occurred, the working memory of the primary and backup commanders needed to be carefully considered to define what information and data could be processed within the mode commander in a transient fashion and which part of the data and information had to be stored so that it could be accessed by the backup commander if needed. While all information could theoretically be stored in a mutually accessible way, the modeling of transient and persistent storage allowed for the assessment of components such as RAM and thus provided inputs to the development of the information architecture and management.

Second, the modeling and simulation of the HotSwap pattern revealed a potential issue with clock frequencies, signal timing, and the respective behavior alignment. Since hardware components are defined according to certain required parameters, such as clock speed and runtime intervals, functions that execute repeatedly, like state checks, were assessed at runtime frequency. During the HotSwap switch from one commander to the other, lost signals were discovered when repeatedly sent signals were sent in the exact moment in which no mode commander was assigned. Due to the inherent functions of SysML, when a signal is sent to a block or state machine, it is either consumed if it can be used by the entity or left un-consumed. In both cases, the signal and its token are used up and cannot be used further without recreating them or triggering a different action with a relaying effect. Although the time spans in which no mode commander is assigned are seemingly insignificant and far smaller than milliseconds, it is possible that signals arrive precisely at the moment the switch occurs, leading to signal loss. While this signal loss can be acceptable for repeated signals, it is potentially critical and or even detrimental for single signals with high importance. This dependence and emergent behavior stemming from hardware parameters was discovered and addressed as a result of the modeling efforts, and two solutions were derived for consideration in the development:

- Option A: Pausing and storing signals for the short period during which the HotSwap occurred, allowing them to be processed afterwards.
- Option B: Implementing signal redundancy in alignment with the runtime and frequency parameters to ensure reception of the signal. This is achieved by adding extra guards in repeated signal senders to check whether the HotSwap is happening concurrently.

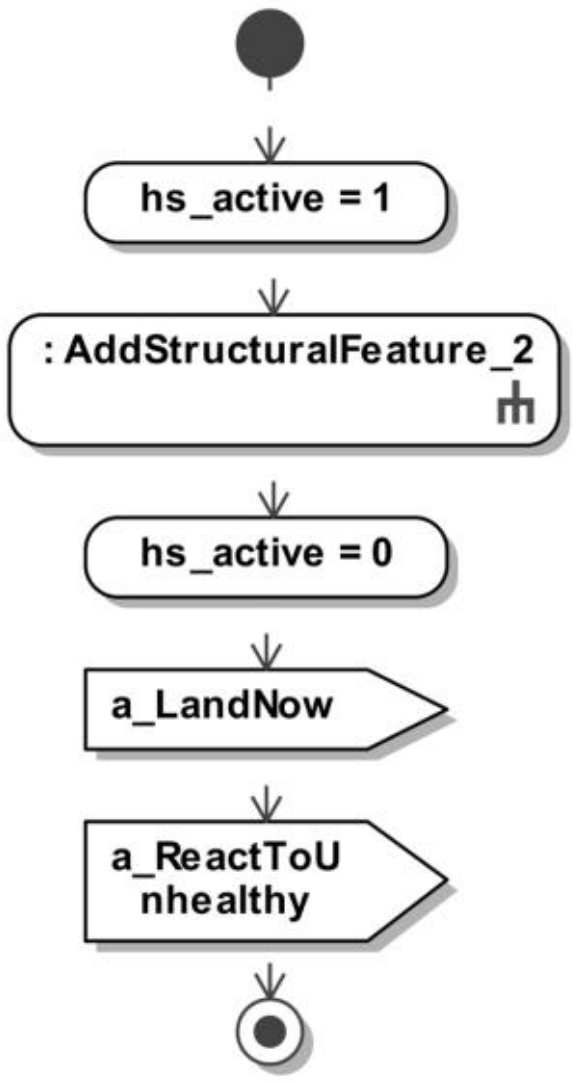

**Fig. 10** Activity setting a parameter when HotSwap is in progress

Option B is chosen to be a more robust approach, as it did not require any interference with the ongoing simulation. To apply this option, a value parameter (*hs_active*) is added to the model to indicate whether a hot swap is happening. This value is then used in the guard conditions to ensure no signal are lost during the transition between mode commanders as depicted in the example in Fig. 9. As shown, the when the hot swap is happening (*hs_active = 1*), the send signal action does not take place, which prevents the loss of any signal. The value of the *hs_active* parameter is set when hot swap is first initiated, before active GNC mode commander is set to any redundant GNC mode commanders as shown in Fig. 10. In this example, not only the value parameters are set to a certain value to indicate that the switch is in progress, but also necessary fault signals are sent to trigger specific behaviors and state changes during a failure after HotSwap is completed.

As the above aspects demonstrate, even seemingly simple actions, such as disconnecting one mode commander and immediately connecting the backup one, can greatly benefit from modeling and behavioral assessment to thoroughly test and define behavior and architecture from the ground up. The assessment of architecture parameters can drive and guide subsequent development steps, while the evaluation and discovery of emergent behaviors within behavioral modeling can reveal potential pitfalls that might otherwise be discovered only after behavior and/or hardware are already defined, or worse, when an error occurs after implementation during operation, with potentially catastrophic consequences. Thus, the case study highlights the value of model-based systems engineering as both an architecture development tool and a simulation/test foundation.

## 5. Code Development From Design

### A. Developer Implementation

Once design has matured to the point that it can be interpreted into a piece of C/C++ code, the design description, such as the state transition diagrams, are provided to a software developer for implementation in code. Traditionally, the developer reads the diagrams and documentation then drafts the code to implement the behavior that the state transition diagram calls for. For instance, when a state has a guard on its exit path, the developer may choose to write an "if" statement in the mode that is TRUE when the parameters/values that match the guard conditions. The next allowed mode transition then resides in the body of the "if" statement.

For SRH, since much of the design already had Ingenuity heritage flight code behind it, the new code required for SRH-specific modes was written directly by a developer interpreting a static copy of the state transition diagrams. The modes are defined as enumerated types in the C/C++ code as 8-bit unsigned integers with a 6-bit mask for a total of 64 modes — which is more than enough for the modes as described in Fig. 4. Event Reports (or EVR for short) associated with the modes and the mode changes were described in a Python script that autogenerated the GNC C/C++ FSW code interactions with the rest of the Avionics FSW. Following the mode and EVR definitions, a callback function associated with each mode is defined and implemented. So while some interface code was auto generated, it was not derived from our executable models.

For the SRH Aerial mode commanders, the GNC flight software was written, tested, and then provided to the Avionics team as a static library. The approach has two main benefits. First, it provides a stardard interface which greatly simplifies intergration with the Avionics FSW. And second, the GNC FSW static library can be easily linked to simulation to verify FSW implementation. In HeliCAT [26]— a simulation framework originally developed for Ingenuity — the GNC FSW static library is extensively used in the closed loop dynamic model for simulation purposes and FSW implementation verfication. Section 6 highlights this approach in mode commander verification.

### B. *Autocoding Executable Models as Flight Software*

Even though SRH did not use the executable models in its GNC FSW development directly, the GNC mode commander structure follows a predictable pattern that lends itself to a potential direct autocoding (at least of stub/boilerplate code) from the executable models. [27] The GNC FSW consists of a main executive loop that, among other things, runs the mode commander executive. At every FC main execution loop, based on the current and next state information, the mode commander executive calls the appropriate GNC mode callback function. This pattern is the same for both the FC and Nav mode commanders and follows the Ingenuity heritage design. [28] Each mode callback method implements 1) the mode transition logic and conditional guards using GNC specific variables and timing information as described by the design intent in Fig. 4, 2) implements call to the GNC guidance code, and 3) implements calls to the estimation, controller and actuation methods. As executable models have more validated syntax implemented, this pattern can be more easily automated from an executable model to generate direct GNC FSW code if desired, at least of the structure of the modes and the allowed exits and guards – the models do not contain real information about the actual algorithms used in the guidance or control, nor do they model the Avionics-based callback functions in F-Prime. Even so, if done right, this approach offers fast, reproducible code directly from the design description (the SysML models), which can significantly reduce human implementation errors. Thus, the SRH team chose to auto-generate FSW code of the mode commanders using the process described in the next section so that it could be qualitatively compared to the existing hand-developed code.

As pointed out in [29], correct implementations of complex hierarchical state machines in flight code are difficult and labor intensive, which drives the interest in autocoding solutions to make the process more reliable and more time efficient. The first JPL mission to use autocoding from state machines was Deep Space 1 in 1998 [29]. Since then, many other missions, including flagships such as the Curiosity and Perseverance Mars rovers, have used similar tools and processes in parts of their FSW implementation [30], [31]. However, these state machines were not fully inherited from the systems engineering models but were rather re-implemented around the initial design in order to meet FSW-specific design requirements. [30]

Multiple tools exist to translate state machines into executable code. An example that has been influential in most JPL efforts at autocoding state machines is the Quantum Platform [32]. The Quantum Platform is a Real-Time Embedded Framework based on Hierarchical State Machines and an event-driven architecture. It provides C/C++ libraries to implement state machine logic, as well as a desktop modeling tool to design state machines in a graphical environment and autocode them into C/C++. For its past missions, JPL has developed a tool internally [33]. The JPL autocoder outputs C/C++ code, leveraging the Quantum Platform libraries for the state machine logic. The motivation for JPL developing their own tool is that teams can ensure that the generated code meets the mission's requirements for coding standards and implementation. There are also other open-source alternatives, such as JPL STARS [34] or Comodo [35], which both have their own mechanics but ultimately output code based on the Quantum Platform as well.

Regardless of which tool a mission selects, the generated code is structured similarly. A series of functions define the states, their possible transitions, and how they respond to events. States can also exhibit behavior on entry, exit and

AutoCode from Executable SysML model

Generated by Comodo (https://github.com/Open-MBEE/Comodo)

```
/**
 * EstModeSpinup state definition
 */
QState nAVMC_NAV MC_unnamedState0_NavModeCommanderState_EstModeSpinup (unnamedState0_NavModeCommanderState *me, QEvt const *e) {
    switch (e->sig) {

        case NAVMC_a_EstSpinupComplete_SIG:
            if (nAVMC_NAV MC_impl_self.get(me->impl, 'active_arbit.arbit_to_heli.active_gnc_mc').in('TakeoffPrep')  || self.get(
            'active_arbit.arbit_to_heli.active_gnc_mc').in('SysID')){
                return Q_TRAN(&nAVMC_NAV MC_unnamedState0_NavModeCommanderState_EstModeIMUAscent);
            }

            return Q_HANDLED();

        case NAVMC_a_EstSpinupFail_SIG:
            return Q_TRAN(&nAVMC_NAV MC_unnamedState0_NavModeCommanderState_EstModeIdle);

    }
    return Q_SUPER(&nAVMC_NAV MC_unnamedState0);
}
```

**Fig. 11** Example raw Autocode output that was compared to hand-developed SRH code for a single estimator mode, with automateable features highlighted

when receiving events, including at a regular interval using Real-Time Interrupts (RTIs). All transition guards as well as all on-entry, on-exit, and on-event behavior specified in the state-machine gets translated into an empty handler function that developers can then hand-code to meet their specific needs. This means that the entire state machine logic is auto generated for the developers, who can focus on the implementation of the mission-specific behaviors and requirements. This process is detailed further in [33]. The results of a Comodo-generated Autocode for a particular estimator mode are shown in Fig. 11. This output was compared to hand-written code and found to have very similar structure and features, with some differences (e.g. switch case implementation vs if/then) being tunable in the autocoder itself.

### *C. Considerations for Autocoding Flight Software*

One limitation of this process is that while SysML is a feature-rich language in the modeling of state machines and their behavior and activities, what can be translated into code is limited to what the translating tool or underlying engine supports. Systems engineers must therefore either limit themselves in their modeling to what can be autocoded, or validation must be performed to confirm that the behavior of the executable model matches that of the implemented software. Furthermore, FSW engineers might need to refine the model before it is autocoded to ensure it is meeting the lower-level FSW design considerations and requirements. [30] For these reasons, should the same model be used for systems design all the way through to FSW autocoding, collaboration between Systems Engineers and FSW engineers is likely preferred early-on in the creation of the state machines.

Another consideration in this approach is how to manage architectural changes after the code has been generated and modified by a developer, If the autocode and hand-written code (say, to add in details on the guidance or control implementation) are deeply intertwined, it may be difficult to integrate/update the autocoded FSW without having to re-do the hand-written code. The SRH team identified this as a concern that, with proper architecting and change control, could be mitigated. When properly implemented, using an autocoder can even facilitate change control, as it is trivial to re-run an autocoder, run git diff on the new code, and highlight specific areas of the code that will need to be changed, rather than requiring a lengthy re-review of every module of hand-written code against a powerpoint slide with the updated architectural diagram. Even so, as SRH inherited a significant amount of hand-developed code that was not structured to make this integration easier, the SRH team did not directly benefit from the use of autocoded FSW.

For future projects, F Prime, the FSW framework used by the SRH Avionics, is in the process of adding state machines to their domain-specific modeling language [21]. Rather than autocoding SysML straight into C++, one could write a translator from SysML to the F Prime modeling language. This may alleviate some of the aforementioned challenges by hiding the FSW implementation details away from the SysML model.

## 6. Verification and Validation of Design

### *A. Test Case Implementation*

Despite being early in the development cycle, the SRH Aerial Mobility team was able to deliver its first drop of GNC FSW, including SRH-specific updates to the mode commanders that matched the developed executable models. As a part of that delivery, the GNC team was responsible for creating unit tests for the new mode commander code. Although SRH had intended to use the executable models to develop a robust traceable verification and validation (V&V) test suite, we were not able to implement a comprehensive set of executable-model-derived test cases before the project ended. Instead, we developed independent tools for V&V and generated three example nominal test cases using the executable models as a proof of concept.

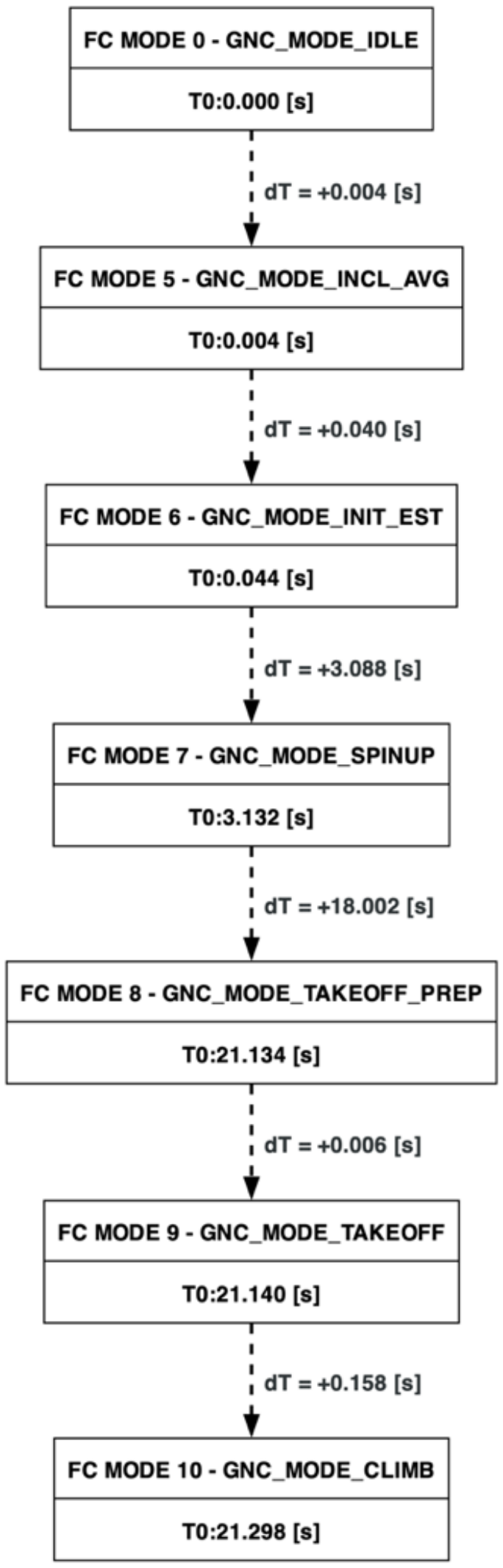

**Fig. 12** Plot for fast mode commander FSW implementation verification; The plot shows the mode transitions that occur during nominal SRH takeoff sequence with timing and FSW specific information; Here, FC is Flight Computer, T0 is the mode start time and dT is the duration; Timing information is displayed in seconds.

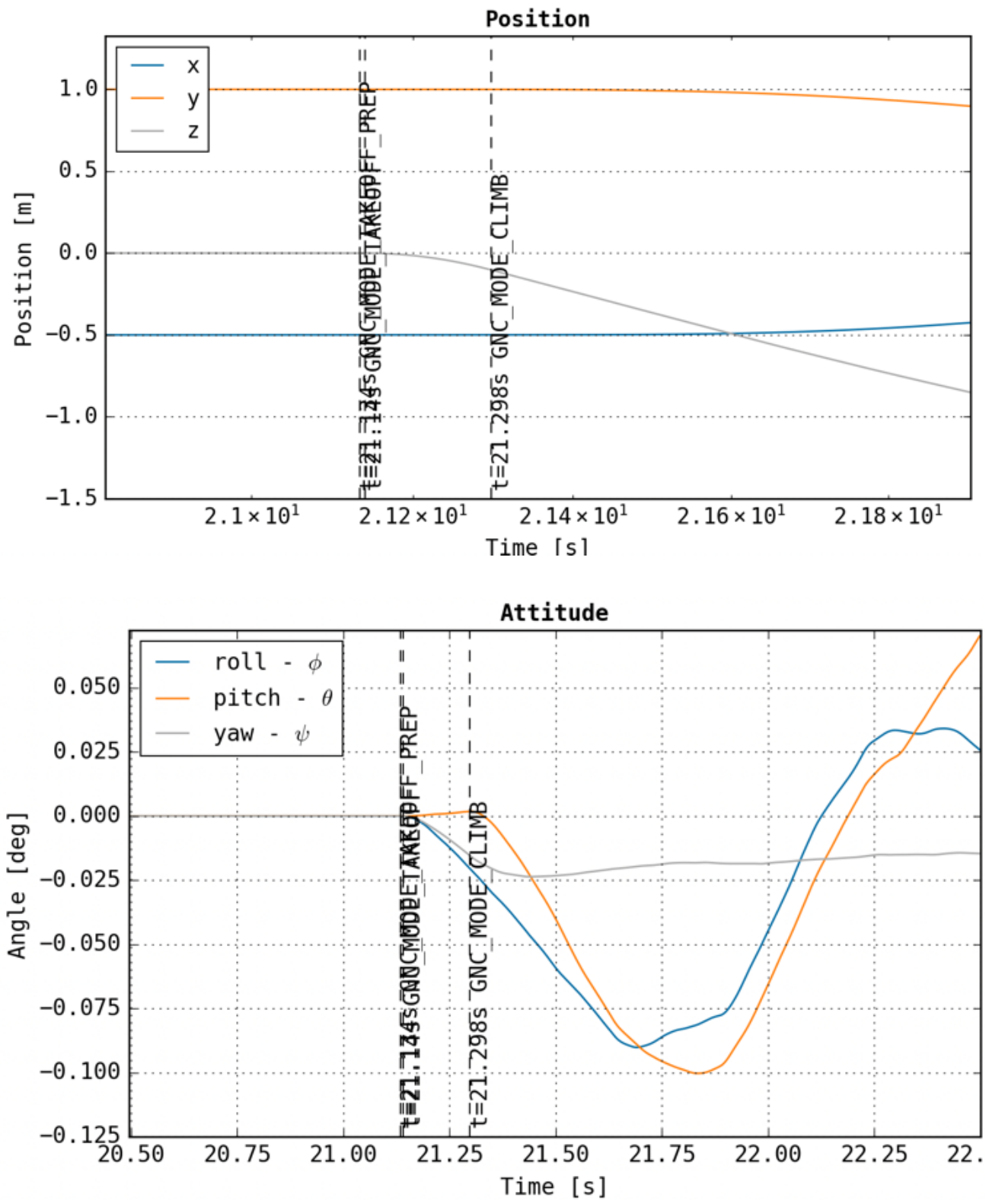

**Fig. 13** Mode changes vs vehicle trajectory

We developed two standalone V&V tests to vet the mode commander: one using a linearized system model to quickly verify and check the mode-commander implementation and a second using a high-fidelity HeliCAT [22] non-linear simulation for end-to-end validation of mode commander implementation against design intent. The fact that these models took as inputs guidance and vehicle trajectories generated by a dynamics model [23] means that they are well-situated to further Monte-Carlo studies. A Python-based tool was written to generate output products such as plots of FSM state-charts and vehicle trajectories. The FSM state-chart portrays the mode commander state-transitions with timing information, mode name and its enumerated integer values. Fig. 12 shows an example of the FSM state chart during a take-off simulation produced by the tool. Fig. 13 shows the corresponding vehicle position and attitude obtained from the takeoff simulation run. We used the information provided in these outputs to evaluate the design intent specifications, particularly transition logic, conditional guards and timing specifications for SRH.

### B. *Executable-Model-Based Test Case Development*

The SRH team also developed a test case based on sequence diagrams that explored the nominal mode transitions using the executable models. Sequence diagrams typically represent the flow of messages or events exchanged between objects, capturing the sequence of interactions that occur between system entities. We defined a nominal egress flight sequence, a nominal non-egress unlocalized flight sequence (such as mapping), and a nominal localized flight sequence (such as a tube retrieval or deposit flight. This definition was initially communicated in excel and was translated into the visual timelines that represent the temporal behavior of the system. Each "swim lane" (or "life line") is an object or actor and each row represents a communication between the system components as a function of time. In SRH, signals can either be coming from *User Control* to represent user commands, or *Automatic Signal Injector* to represent signals generated from the onboard sensors or actions. An example of the test scenario captured in SysML is shown in Fig. 14.

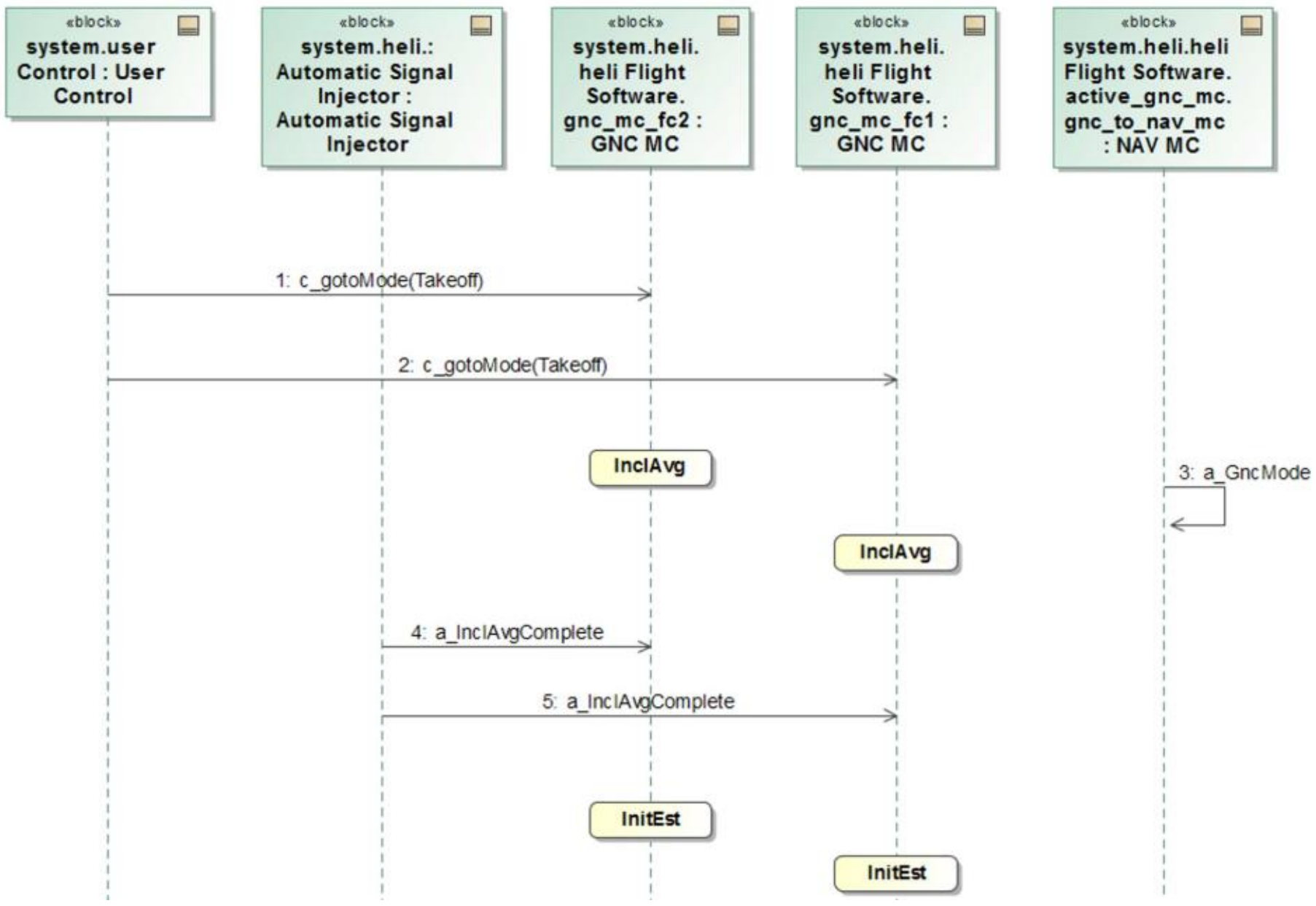

**Fig. 14** Example test scenario captured in SysML

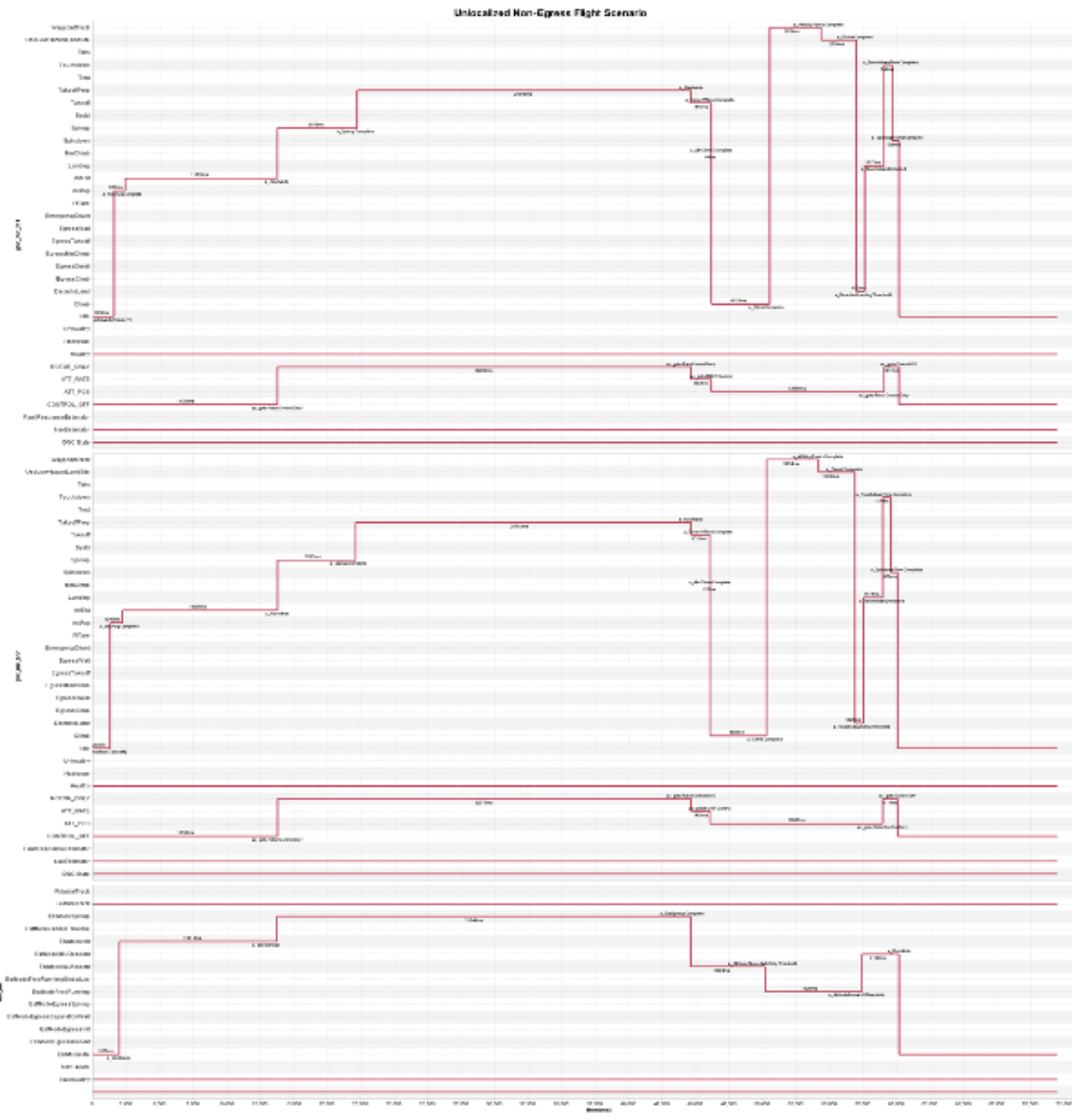

**Fig. 15** State timeline chart for mode commanders

As signals trigger state transitions, SysML state machines that are linked to the sequence diagrams execute, providing a dynamic and automated representation of the system's behavior. This kind of diagram also allows the user to set timing ranges, limits, and constraints of the signals or modes, allowing a more complete examination of timing-related interactions in the model. The resulting state timeline chart in Fig. 15 is generated by executing this flight scenario via simulation configurations in Cameo Systems Modeler. In this figure, the active states of the mode commanders during the simulation can be depicted. Explicit checks can be placed to ensure that the system is in the correct state at each point in the sequence, and timing constraints can be added to provide a more realistic test scenario. By mapping signal exchanges and state transitions into formal models, the diagrams effectively serve as executable test cases that simulate system behavior against specified requirements or scenarios. These exchanges can be refined in the simulation then exported/processed into test code that can in turn be used to test the final FSW code in a tighter coupling of the V&V from design to implementation. Even on a truncated timeline, SRH was able to show that this method of generating test cases has great potential to improve subsystem level V&V and expedite the creation of meaningful unit tests that can carry forward into testbed development and beyond.

## 7. Conclusions

SRH chose to design its updates to the Aerial mode commanders using formal executable models in SysML with the goal of examining what benefits the formalism might offer to subsystem-level software design. We discovered that the mature implementation approach – facilitated by the Ingenuity heritage we were able to leverage – helped accelerate the fidelity of the models, leading us to produce a set of valuable modeling patterns for complex but common mode commander features such as interdependent state machines and hot swap modeling. Further, the use of executable, formalized models had a number of realized benefits to the design and development team. In particular, it helped to enforce a rigor of definition and syntactic accuracy in the transitions and definitions that could have otherwise been ambiguous if described in a less exacting tool (like powerpoint). It also helped systems engineers identify early issues with the architecture and facilitate a smooth transition from design to implementation. The approach provided a clear way to preserve design architecture intent in the code development by forming the architecture definition into a form more directly translatable to code, and while SRH did not use the autocoding capabilities of the technique for writing its flight code, it did prototype this approach and found that it had promise in some scenarios so long as change control between autocoded elements and hand-written elements was managed carefully. Finally, while many benefits to V&V would have been borne out in later project phases, the initial set of test cases suggested that this approach offers perhaps a more systematic and explicit method of building test suites that can be automated and thorough, complementing the more end-to-end simulations that are used to ensure proper system level behavior overall. The investment in developing a working simulation of the architecture and conceptual design was well worth the benefits even at the post-PDR level we achieved. Traditional ways to achieve this level of communication and consistency across the design-development continuum – design specifications, behavior descriptions, testing matrices etc. – all work well but are subject to the ambiguities of natural language and human error. It can also be time-consuming to translate from design to code. We found the executable models facilitated team interactions and approaches that were more likely to lead to a more robust, efficient, and reliable product. Furthermore, the model's extensibility allows for increased fidelity and the incorporation of additional test cases, suggesting that this approach to systems definition has the ability to grow and mature throughout the project lifetime.

Acknowledgment

The authors would like to thank a number of people for their engineering expertise that made this work possible: Robert Karban, Dylan Conway, Johnny Lam, Dan Burkhardt, and the entire Sample Recovery Helicopter Aerial Mobility GNC code development team. The research was carried out at the Jet Propulsion Laboratory, California Institute of Technology, under a contract with the National Aeronautics and Space Administration (80NM0018D0004).

## References

[1] Hirshorn, Steven; Voss, Linda; Bromley, Linda, "NASA Systems Engineering Handbook," 2017.

[2] A. Reichwein and C. Paredis, "Overview of Architecture Frameworks and Modeling Languages," in *Proceedings of the ASME 2011 International Design Engineering Technical Conferences &*

*Computers and Information in Engineering Conference*, Washington, DC, USA, August 29-31, 2011.

[3] S. Zhong, A. Scarinci and A. Cicirello, "Natural Language Processing for systems engineering: Automatic generation of Systems Modelling Language diagrams," *Knowledge-Based Systems,* vol. 259, no. 110071, 10 January 2023.

[4] Friedenthal, Sanford; Moore, Alan; Steiner, Rick, A Practical Guide to SysteML The Systems Modeling Language, Boston: Elsevier, 2015.

[5] T. Weilkins, J. Lamm, S. Roth and M. Walker, Model-Based System Architecture, Hoboken, NJ: Wiley, 2022.

[6] N. J. A. E. Trawny, E. Bailey, G. Massone, M. Reid, T. P. Setterfield, Y. Cheng, G. Sellar and J. Soto, "The Enhanced Lander Vision System for Mars Sample Retrieval Lander Entry Descent and Landing", not Mars Sample Return Lander," in *AIAA SCITECH 2024 Forum*, Orlando, FL, 2024.

[7] R. Karban, N. Jankevičius and M. Elaasar, "ESEM: Automated Systems Analysis using Executable SysML Modeling Patterns," *INCOSE International Symposium,* vol. 26, no. 1, pp. 1-24, 13 September 2016.

[8] J. Dahmann, A. Markina-Khusid, A. Doren, T. Wheeler, M. Cotter and M. Kelly, "SysML executable systems of system architecture definition: A working example," in *2017 Annual IEEE Systems Conference*, Montreal, QC, Canada, 2017.

[9] R. Karban, F. Dekens, S. Herzig, M. Elaasar and N. Jankevičius, "Creating system engineering products with executable models in a model-based engineering environment," in *SPIE 9911, Modeling, Systems Engineering, and Project Management for Astronomy VII*, Edinburgh, United Kingdom, 2016.

[10] G. Kminek, M. A. Meyer, D. W. Beaty, B. L. Carrier, T. Haltigin and L. E. Hays, "Mars Sample Return (MSR): Planning for Returned Sample Science," *Astrobiology,* vol. 22, no. S1, 2022.

[11] B. K. Muirhead, A. K. Nicholas, J. Umland, O. Sutherland and S. Vijendran, "Mars Sample Return Campaign Concept Status," *Acta Astronautica,* vol. 176, pp. 131-138, November 2020.

[12] Mier-Hicks, Fernando, Håvard Fjær Grip, Arash Kalantari, Scott J. Moreland, Benjamin Pipenberg, Matthew Keennon, Timothy K. Canham, Michael T. Pauken, Emmanuel Decrossas, Theodore Tzanetos and J. Balaram., "Sample Recovery Helicopter," in *2023 IEEE Aerospace Conference*, 2023.

[13] Tzanetos, T ., Aung, M.M., Balaram, J., Grip, H., Karras, J.T., Canham, T.K., Kubiak, G., Anderson, J.L., Merewether, G.B., Starch, M.D., Pauken, M.T., Cappucci, S., Chase, M., Golombek, M.P., Toupet, O., Smart, M.C., Dawson, S.F., Ramirez, E.B., Lam, J.S, "Ingenuity Mars Helicopter: From Technology Demonstration to Extraterrestrial Scout.," in *2022 IEEE Aerospace Conference*, 2022.

[14] Grip, H.F., Johnson, W., Malpica, C., Scharf, D.P., Mandic, M., Young, L., Allan, B., Mettler, B. and San Martin, M., "Flight dynamics of a Mars Helicopter," NASA, 2017.

[15] Pipenberg, B., Keennon, M., Tyler, J.D., Hibbs, B.D., Langberg, S.A., Balaram, J., Grip, H.F., & Pempejian, J., "Design and Fabrication of the Mars Helicopter Rotor, Airframe, and Landing Gear Systems," *AIAA Scitech 2019 Forum,* 2019.

[16] Pipenberg, Benjamin, Matthew Keennon, Sara A. Langberg and Jeremy D. Tyler., "Development of the Mars Helicopter Rotor System," in *Proceedings of the Vertical Flight Society 75th Annual Forum*, 2019.

[17] Schatzman, Natasha, Athena Chan, Michael Fillman, Larry Meyn, Vinod Gehlot, Kenneth Glazebrook, Diego Santillan, and Paulina Ridland., "Performance Analysis and Data Processing for the Mars Sample Recovery Helicopter in the Jet Propulsion Laboratory 25-ft Space Simulator," in *80th Vertical Flight Society's (VFS) Annual Forum & Technology Display*, 2024.

[18] Withrow-Maser, Shannah et al. , "Mars Sample Recovery Helicopter: Rotorcraft to Retrieve the First Samples from the Martian Surface," in *Vertical Flight Society 79th Annual Forum & Technology Display*, 2023.

[19] Balaram, J. & Aung, MiMi & Golombek, Matthew., "The Ingenuity Helicopter on the Perseverance Rover.," *Space Science Reviews,* vol. 217, no. 4, 2021.

[20] D. S. Bayard, D. T. Conway, R. Brockers, J. Delaune, L. Matthies, H. Grip, G. Merewether, T. Brown and A. M. San Martin, "Vision-Based Navigation for the NASA Mars Helicopter," in *AIAA SciTech Forum*, San Diego, California, 7-11 January 2019.

[21] R. L. J. T. K. Bocchino, G. J. Watney, L. J. Reder and J. W. Levison, "F Prime: An Open-Source Framework for Small-Scale Flight Software Systems," in *32nd Annual AIAA/USU Conference on Small Satellites*, Logan, UT, 2018.

[22] C. Leake, H. Grip, V. Steyert, T. D. Hasseler, M. Cacan and A. Jain, "HeliCAT-DARTS: A High Fidelity, Closed-Loop Rotorcraft Simulator for Planetary Exploration," *IEEE Aerospace,* vol. 11, no. 727, 2024.

[23] H. Grip, J. Lam, D. Bayard, D. Conway, G. Singh, R. Brockers, J. Delaune, L. Matthies, C. Malpica, T. Brown, A. Jain, M. San Martin and G. Merewether, "Flight Control System for NASA's Mars Helicopter," in *AIAA SciTech Forum*, San Diego, California, 7-11 January 2019.

[24] "ALH APIs," No Magic Inc., 2024. [Online]. Available: https://docs.nomagic.com/display/MSI2022xR2/ALH+APIs. [Accessed 13 September 2024].

[25] "Cameo Systems Modeler," Dassault Systèmes, [Online]. Available: https://www.3ds.com/products/catia/no-magic/cameo-systems-modeler. [Accessed 13 September 2024].

[26] C. Leake, H. Grip, V. Steyert, T. Hasseler, M. Cacan and A. Jain, "HeliCAT-DARTS: A High Fidelity, Closed-Loop Rotorcraft Simulator for Planetary Exploration," vol. 11, no. 9, 2024.

[27] H. Grip, J. Lam, D. S. Bayard, C. T. Dylan, G. Singh, B. Roland, L. H. Matthies, C. Malpica, T. L. Brown and A. A. Jain, "Flight Control System for NASA's Mars Helicopter," in *AIAA SciTech 2019*, San Diego, California, 2019.

[28] T. Canham, "The Mars Ingenuity Helicopter - A Victory for Open-Source Software," in *IEEE Aerospace Conference (AERO)*, Big Sky, Montana, 2022.

[29] E. Benowitz, K. Clark and G. J. Watney, "Auto-coding UML statecharts for flight software," in *Space Mission Challenges for Information Technology* , Pasadena, CA, 2006.

[30] E. Benowitz, "UML Statechart Autocoding for the Mars Science Lab (MSL) Mission," in *2012 Workshop on Spacecraft Flight Software*, San Antonio, TX, November 7-9, 2012.

[31] S. Scandore, Interviewee, *Hierarchical State Machines Persevere on Mars.* [Interview]. 25 May 2021.

[32] M. Samek, Practical Statecharts in C/C++, San Francisco: CMP Books, 2002.

[33] K. L. Wagstaff, E. B. D. J. Benowitz, K. Peters and G. Watney, "Automatic Code Generation for Instrument Flight Software," in *Meeting: 9th International Symposium on Artificial Intelligence, Robotics, and Automation in Space* , Universal City, CA, 2008.

[34] "JPLOpenSource STARS," JPL, [Online]. Available: https://github.com/JPLOpenSource/STARS. [Accessed 16 September 2024].

[35] "Open-MBEE Comodo," JPL, [Online]. Available: https://github.com/Open-MBEE/Comodo. [Accessed 16 September 2024].

[36] N. M. Inc., "Manipulating runtime objects using scripts," [Online]. Available: https://docs.nomagic.com/display/MSI2024xR1/Manipulating+runtime+objects+using+scripts. [Accessed 13 September 2024].

[37] Johnson, W., Withrow-Maser, S., Young, L., Malpica, C., Koning, W.J., Kuang, W., Fehler, M., Tuano, A., Chan, A., Datta, A. and Chi, C, "Mars Science Helicopter Conceptual Design," NASA, 2020.

[38] R. L. Bocchino Jr., T. K. Canham, G. J. Watney, L. J. Reder and J. W. Levison, "F Prime: An Open-Source Framework for Small-Scale Flight Software Systems," in *Proceedings of the Small Satellite Conference, Advanced Technologies 2*, 2018.

[39] R. Bocchino, T. Canham, G. Watney, L. Reder and J. Levison, ""F Prime: An Open-Source Framework for Small-Scale Flight Software Systems," in *32nd Annual Small Satellite Conference*, Logan, UT, 2018.

## BIOGRAPHY

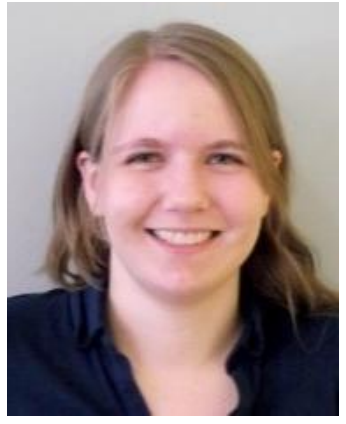

***Laura Jones-Wilson** obtained her Ph.D. at Cornell University in Aerospace Engineering specializing in the Dynamics and Controls of space systems. She then joined JPL in 2012 as a guidance and control systems engineer taking on roles including the PI on a Mars Sample Capture technology development effort and the co-manager of the SmallSat Dynamics Testbed. She joined Europa Clipper in 2015 as the instrument engineer for UVS then REASON, and later served as the lead systems engineer for the REASON-Solar Array Integrated Product Team and Payload V&V Lead through instrument deliveries to ATLO. She then became the lead systems engineer for the Aerial Mobility team of the Sample Recovery Helicopter (SRH) and now serves as the Aero and Flight Dynamics Lead for Chopper.*

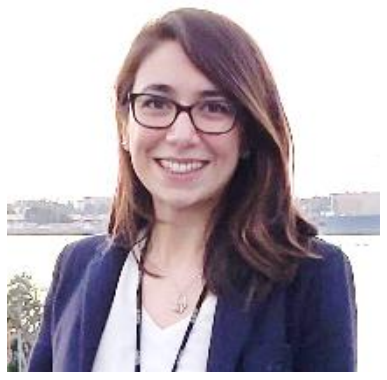

***Dr. Fatma Karagoz** is a Systems Engineer at NASA Jet Propulsion Laboratory. She holds an M.S. and Ph.D. in Aerospace Engineering from Georgia Institute of Technology. She is currently leading a systems development environment as a member of the Systems Products Solutions Engineering group. Her primary research areas include multi-disciplinary design and optimization, model-based systems engineering, complex system architecting, and product family and platform design. She is currently involved in the Mars Sample Return project at JPL.*

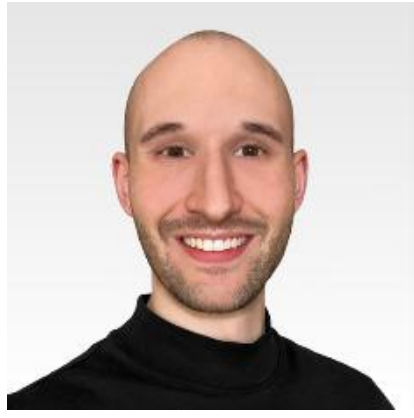

***Maximilian Vierlboeck** is a Systems Engineer at NASA's Jet Propulsion Laboratory, where he focuses on the development of advanced systems engineering approaches and digital engineering frameworks. As the product owner of the requirements engineering suite, his work involves leading the design, implementation, and integration of innovative technologies that streamline both requirements and systems engineering across missions and projects. Maximilian holds a Ph.D. from Stevens Institute of Technology and previously earned bachelor's and master's degrees in mechanical engineering from the Technical University of Munich (TUM). Before his time* *at JPL, he gained practical experience in product development with Knorr-Bremse. Maximilian's research has included topics such as complexity assessment frameworks, natural language processing for requirements engineering, and agile methodologies. He has been recognized with multiple awards for his research excellence, publications, and academic achievements.*

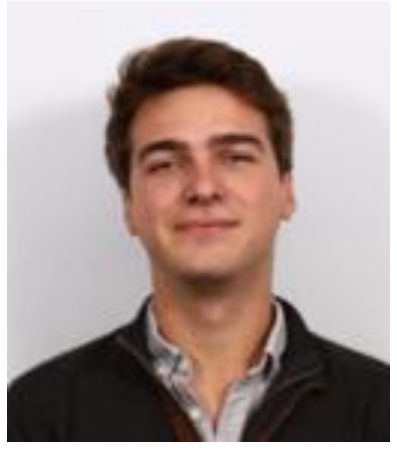

***Thomas Boyer-Chammard** has a M.Eng in Computer Science from École Centrale de Lille, France. Thomas joined JPL in 2019 as a Software Systems Engineer during his Master's. His early efforts focused on enhancing Systems Engineering processes with software solutions. Thomas worked with SysML v2 and researched how to use natural language processing to augment models. Later, Thomas became the Product Owner for SEED, developing processes for embedded software development at JPL. Today, Thomas serves as the Deputy Cognizant Engineer for F Prime, an open-source Flight Software framework.*

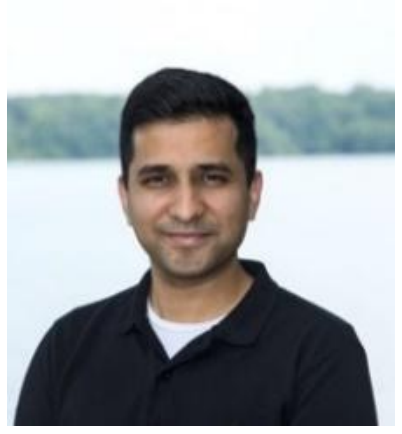

***Dr. Vinod Gehlot** is a Robotics System Engineer in the Guidance and Control Section at NASA/Caltech Jet Propulsion Laboratory in Pasadena, CA. He received his Bachelor of Science and Master of Science degrees in aerospace engineering from Embry-Riddle Aeronautical University in Daytona Beach Fl, and his Ph.D. from the University of Tennessee, Knoxville. Prior to JPL, Vinod was a Research Engineer in the Mechanical Engineering Department at Texas A&M University, College Station, TX. As a researcher, he is passionate about dynamical systems, adaptive systems, control theory and multi-agent robotics. At JPL, he his helping in developing technologies for future rotorcraft systems at Mars and GNC for the Habitable Worlds Observatory. Before pursuing his doctorate, he worked at the Eagle Flight Research Center at Embry-Riddle Aeronautical University, where he designed and developed the hardware, software, and algorithms for the autonomous flight control of aerial robotic vehicles with highly articulated rotors. Vinod has authored two patents three patents (two issued) on mechanical mechanisms for the counter-rotation of axial compressors in gas turbine engines. Besides his academic and research endeavors, Dr. Gehlot also has extensive experience in the development of high-performance embedded hardware and software, including field-programmable gate arrays (FPGAs) for the test and development of advanced flight control and sensor fusion algorithms.*